\documentclass[runningheads]{llncs}

\usepackage{graphicx}
\usepackage[utf8]{inputenc}
\usepackage[T1]{fontenc}

\usepackage{mathtools}	
\usepackage{varwidth}
\usepackage{adjustbox}
\usepackage{multirow}
\usepackage{amsmath}

\usepackage{algpseudocode,algorithm,algorithmicx}

\usepackage{booktabs}
\usepackage{multirow}
\usepackage{tabularx}

\title{Discovering Sequential Patterns in Event-Based Spatio-Temporal Data by Means of Microclustering - Extended Report}

\titlerunning{Sequential Patterns in Event-Based Spatio-Temporal Data}        

\author{Piotr S. Maciąg
}

\institute{
	Institute of Computer Science, Warsaw University of Technology,\\
	Nowowiejska 15/19,\\
	00-665, Warsaw, Poland\\
	\email{pmaciag@ii.pw.edu.pl}          
}

\begin{document}

\maketitle

\begin{abstract}
	
In the paper, we consider the problem of discovering sequential patterns from event-based spatio-temporal data. The problem is defined as follows: for a set of event types $F$ and for a dataset of events instances $D$ (where each instance in $D$ denotes an occurrence of a particular event type in considered spatio-temporal space), discover all sequential patterns defining the following relation between any event types participating in a particular pattern. The following relation $\rightarrow$ between any two event types, denotes the fact that instances of the first event type attract in their spatial proximity and in considered temporal window afterwards occurrences of instances of the second event type. In the article, the notion of sequential pattern in event-based spatio-temporal data has been defined and the already proposed approach to discovering sequential pattern has been reformulated. We show, it is possible to efficiently and effectively discover sequential patterns in event-based spatio-temporal data.  


\end{abstract}

\section{Introduction}
\label{Sec:Intro}

Discovering knowledge from spatio-temporal data is gaining attention of researchers nowadays. Based on literature we can distinguish two basic types of spatio-temporal data: event-based and trajectory-based \cite{ref1284:Li2014}. The latter refers to the situation where for a set of objects and their movements (trajectories), we would like to find several types of patterns: clustering trajectories \cite{ref1284:Nanni2006}, discovering representative trajectory \cite{ref1284:Lee2007}, finding frequently visited areas \cite{ref1284:Mamoulis2004}. We do not consider this type of data in our paper. The appropriate review of frequent patterns mining methods for this type of data may be found in \cite{ref1284:Li2014}.

\begin{figure}[h!t]
	\includegraphics[width=1.0\linewidth]{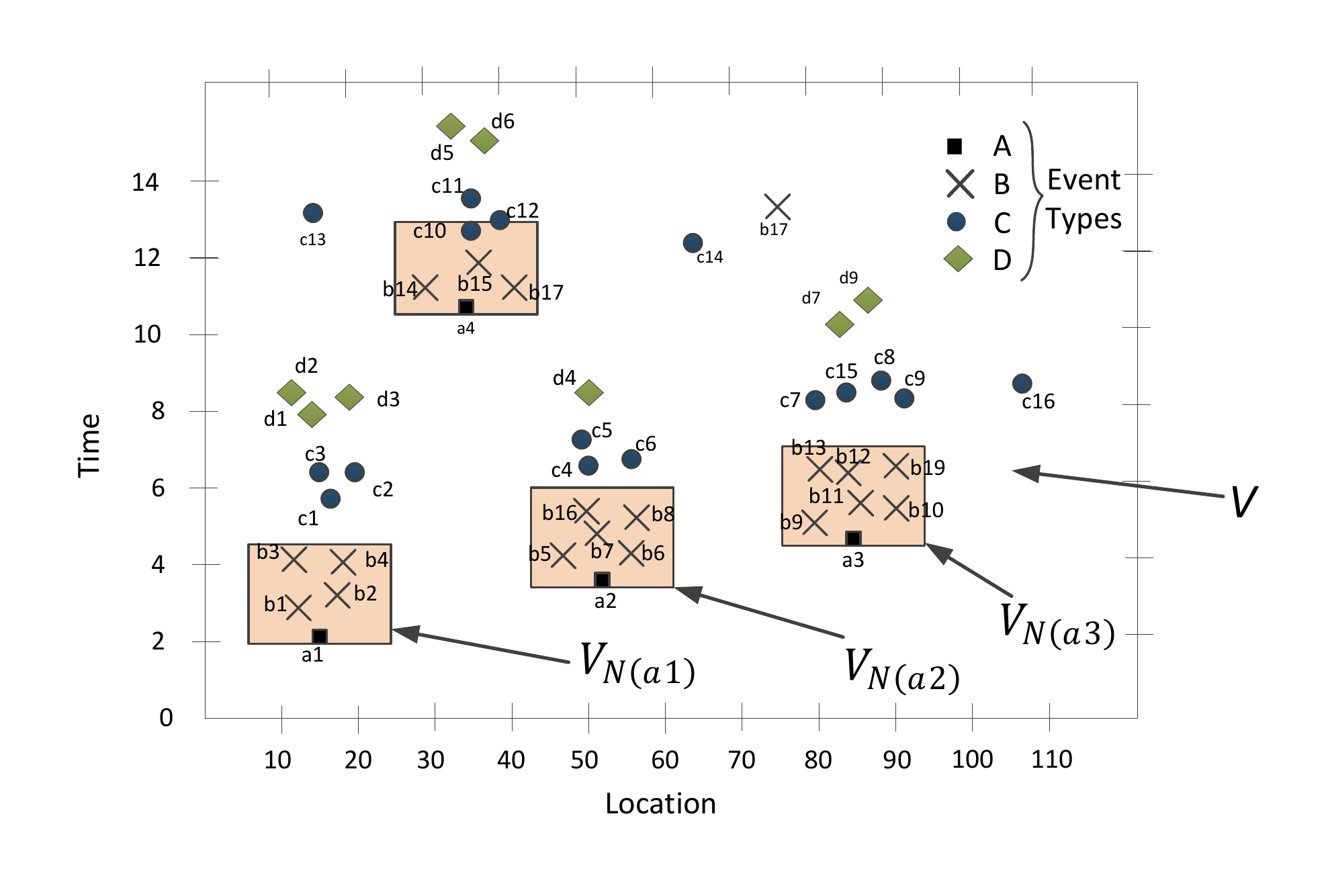}
	\caption{Visualization of the dataset from Table \ref{Table:ExamDataset}.}
	\label{Fig:1}      
\end{figure}

On the other hand, event-based spatio-temporal data is described by a set of event types $F = \{f_1, f_2, $ $\dots, f_n\}$ and a set of instances $D$. Each instance $e \in D$ denotes an occurrence of a particular event type from $F$ and is associated with instance identifier, location in spatial dimension and occurrence time. In a more complex scenario, it is possible that an instance is also associated with a particular numerical value. For example, for an event type \textit{temperature}, an instance of this event type may contain a value $20^{\circ}C$. However, we do not consider this situation in our paper. Table~\ref{Table:ExamDataset} provides possible sets $D$ and $F$. The dataset given in Table~\ref{Table:ExamDataset} is depicted in Fig.~\ref{Fig:1}. Event-based spatio-temporal data and the problem of discovering frequent sequential patterns in this type of data have been introduced in \cite{ref1284:Huang2008}.

\begin{table}[h!t]
	\centering
	\caption{An example of a spatio-temporal event-based dataset.}
	\label{Table:ExamDataset}       
	\begin{tabular}{llll}
		\hline\noalign{\smallskip}
		Instance identifier & Event type & Spatial location & Occurrence time  \\
		\noalign{\smallskip}\hline\noalign{\smallskip}
		a1 & A & 19 & 1 \\
		a2 & A & 83 & 1 \\
		\vdots & \vdots & \vdots & \vdots \\
		b1 & B & 25 & 3 \\
		b2 & B & 1 & 3 \\
		\vdots & \vdots & \vdots & \vdots \\
		c1 & C & 25 & 7 \\
		c2 & C & 15 & 7 \\
		\vdots & \vdots & \vdots & \vdots \\
		d1 & D & 21 & 11 \\
		d2 & D & 13 & 12 \\
		\vdots & \vdots &  \vdots & \vdots \\
		\noalign{\smallskip}\hline
	\end{tabular}
\end{table}

Spatio-temporal data occurs frequently in real world. Possible applications considered in literature are as follows. \cite{ref1284:Wang2013} considers the problem of discovering ozone polluted areas. For a set of measurements containing informations about ozone pollution in Houston city, the task is to find polygons denoting areas with similar ozone pollution. Additionally, each measurement contains occurrence time, so the applied algorithm may focus on analyzing changes in ozone pollutions. The similar task is considered in \cite{ref1284:Wang2014} and \cite{ref1284:Iyengar2004}. 

Discovering knowledge from spatial data is a widely known topic in the literature. Among the most important problems in that area is discovering spatial collocation or association rules. According to \cite{ref1284:Huang2004}, for a set of predefined event types a spatial collocation is defined as an occurrence of instances of particular event types in spatial proximity. The effective algorithms for discovering spatial collocations are provided in \cite{ref1284:Bembenik2009} and \cite{ref1284:Boinski2014}. Clustering problem in spatial data is widely considered in literature \cite{ref1284:Ester1996}, \cite{ref1284:Zhang1996}. The worth mentioning problem is clustering complex spatial data (for example polygons or areas). The problem has been introduced in \cite{ref1284:Ng2002}, where the authors proposes heuristics for measuring distance between polygons and provide appropriate clustering algorithm - CLARANS. The problem of developing appropriate distance measures and clustering algorithms for complex spatial object should be addressed in future publications. 

The task of mining spatio-temporal sequential patterns in given datasets $F$ and $D$ may be defined as follows. We assume that the \textit{following} relation (or attraction relation) $f_{i_1} \rightarrow f_{i_2}$ between any two event types denotes the fact, that instances of type $f_{i_1}$ attract in their spatial and temporal neighborhoods occurrences of instances of type $f_{i_2}$. The strength of the following relation $f_{i_1} \rightarrow f_{i_2}$ is investigated by comparing the density of instances of type $f_{i_2}$ in spatio-temporal neighborhoods of instances of type $f_{i_1}$ and the density of instances of type $f_{i_2}$ in the whole dataset $D$ and spatio-temporal embedding space $V$. We provide the strict definition of density in Section \ref{Sec:Basic notions}. The problem is to discover all significant sequential patterns defined in the form $f_{i_1} \rightarrow f_{i_2} \rightarrow \dots \rightarrow f_{i_k}$. We assume that in a given pattern the following relation $f_{i_1} \rightarrow f_{i_2}$ is preserved between any consecutive event types participating in that pattern. 

The layout of the article is as follows. In Section~\ref{Sec:RelatedWorks} we provide recall some articles important in the field of discovering patterns in transaction databases. In Section~\ref{Sec:Basic notions} we define elementary notions. We provide our modifications and the Micro-ST-Miner algorithm in Section~\ref{Sec:MircoClust}. The results of our experiments are given in Section~\ref{Sec:ExperimentalResults}. 

\section{Related Works}
\label{Sec:RelatedWorks}

In this section, we provide a review of topics related to the problem considered in our publication. In particular, we consider the problem of discovering frequent patterns as a data mining task. The problem considered in the publication is related to the frequent patterns discovery problem. There exist a broad literature considering topic of discovering frequent patterns in transaction databases (that is, databases containing transaction of purchased items). The problem has been originally formulated in \cite{ref1284:Agrawal1993}.

The most famous algorithms in the considered field are Apriori for generating frequent itemsets and Apriori-Gen for generating association rules \cite{ref1284:Agrawal1994}. Apriori generates a set of frequent patterns using two phases: first, based on the set of frequent $k$-length patterns (that is, patterns containing $k$ items), the set of $k+1$-length candidate patterns is generated. In the second phase, the frequency (that is, the number of occurrences in a dataset) of each candidate is verified and non-frequent candidates are pruned (that is, candidate patterns with frequency below given threshold). Additionally, the candidates are pruned by means of the fact, that any candidate containing a non-frequent subpattern cannot be among frequent patterns (this property is obtained from the problem formulation and in literature is known as anti-monotone \textit{apriori prune rule}). Based on the set of already discovered frequent patterns, the algorithm may proceed with generating $k+2$-length patterns and so forth. The algorithm stops, when the candidate set is empty. 

The authors of \cite{ref1284:Han2004} proposes an algorithm for discovering frequent patterns by means of the structure FrequentPattern-tree (FP-tree). The basic idea of the algorithm presented in \cite{ref1284:Han2004} is to efficiently generate frequent patterns omitting the phase of candidate generation (the phase used in Apriori algorithm). The FP-tree is created as follows: in the first phase, a scan of a database of transactions is performed and all frequent single items are found. In the second step, the database is scanned again, a sequence of frequent single items is identified in each transaction and, if the sequence is not empty, the corresponding path (starting from the root of the tree) is inserted or updated in the already created FP-tree. The nodes in the created or updated path correspond to single items occurring in the sequence.  

The problem of discovering sequential patterns in transaction databases has been formulated in \cite{ref1284:Agrawal1995}. In this context, a sequential pattern is defined as follows. For a set of customers and having a list of transactions for each customer, assume that each transaction contains a set of items. A sequential pattern is a list of sets of items. Each set of items in the list has to be contained in a particular transaction of a customer in the following manner: transaction containing the second set should occur after the transaction containing the first set, transaction containing the third set should occur after the transaction containing the second set and so forth. The number of customers to which the pattern may be matched is defined as the support of the pattern. In \cite{ref1284:Agrawal1995}, the authors provide the following example of a sequential pattern: ''5\% of customers bought 'Foundation' and 'Ringworld' books in one transaction, followed by 'Second Foundation' book in a later transaction'', for a database containing customers and books purchased in transactions. This problem has been extended in \cite{ref1284:Srikant1996}, where the authors introduced the notion of generalized sequential pattern and provided an algorithm GSP for discovering patterns of that type. The notion of sequential pattern is extended to the notion of generalized sequential pattern by introducing temporal constraints for sets of items (e.g. time between the first and the last sets of items in a given pattern must not exceed given interval).

\section{Basic Notions}
\label{Sec:Basic notions}

Let us consider the spatiotemporal event-based dataset given in Fig.~\ref{Fig:1}. The dataset contains $F = \{A, B, C, D\}$ event types and $D = \{a1, a2, \dots, d9\}$ events instances. The dataset is embedded in a spatiotemporal space $V$, by $|V|$ we denote the volume of that space, calculated as the product of spatial area and size of time dimension. Spatial and temporal sizes of spatiotemporal space are usually specified by the domain of considered task (e.g. for a weather dataset which may be used for discovering patterns between climate event types, the spatial area is specified by geographical coordinates of registered instances and size of time dimension is given by times of occurrences of instances). On the other hand, by $V_{N(e)}$ we denote the spatiotemporal neighborhood space of instance $e$ (the formal statement of $ V_{N(e)} $ is given in Definition~\ref{Def:NeighborhoodSpace}). Sizes and shape of $V_{N(e)}$ are usually given by an expert.

\begin{definition}(Neighborhood space)
	By $V_{N(e)}$ we denote the neighborhood space of instance $e$. $\mathcal{R}$ denotes spatial radius and $\mathcal{T}$ temporal interval of that space (and, due to that, $ V_{N(e)} $ has a cylindrical shape). 
	\label{Def:NeighborhoodSpace}
\end{definition}

\begin{figure}[h!t]
	\centering
	\includegraphics[width=1.0\linewidth]{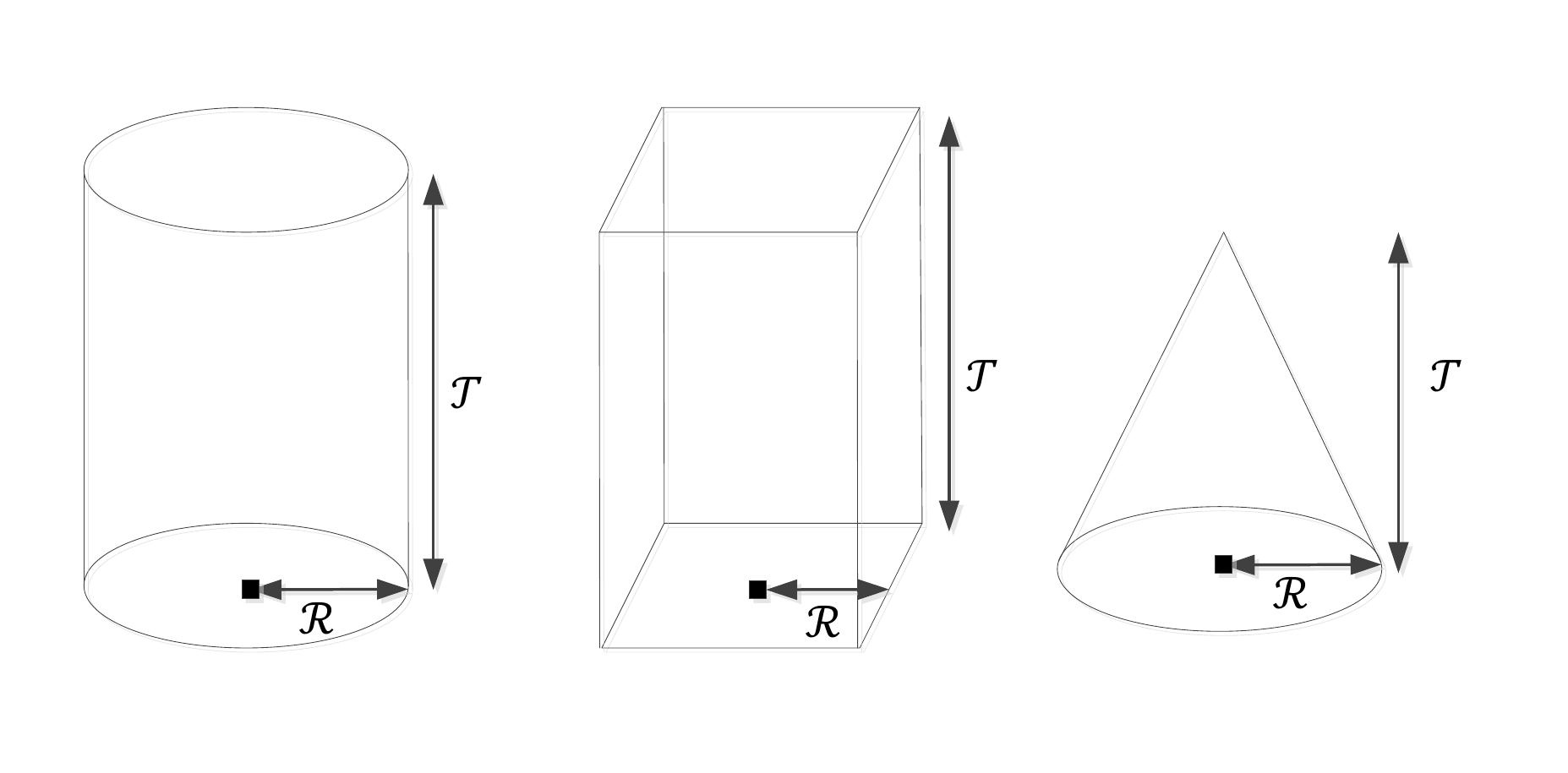}
	\caption[]{Examples of types of neighborhood spaces $V_{N(e)}$ (figure adapted from \cite{ref1284:Huang2008})}
	\label{Fig:2}
\end{figure}

Let us comment Definition~\ref{Def:NeighborhoodSpace} and let consider Fig.~\ref{Fig:2} where we denote possible shapes of neighborhood spaces. It is possible to select other type of neighborhood space than provided in Definition~\ref{Def:NeighborhoodSpace}. For example, it may be appropriate to select conical neighborhood space for some applications.

\begin{definition}(Neighborhood definition \cite{ref1284:Huang2008})
	For a given occurrence of event instance $e$, the neighborhood of $e$ is defined as follows:
	\begin{equation}
	\begin{split}
	N(e) = & \{p | p \in D \land distance(e.location, p.location) \leq \mathcal{R} \\ 
	& \land (p.time - e.time) \leq \mathcal{T} \land p.type = e.type\}
	\end{split}
	\end{equation}
	where $\mathcal{R}$ denotes spatial radius and $\mathcal{T}$ temporal interval of the neighborhood space $V_{N(e)}$.
	\label{Def:Neighborhood}
\end{definition}

The neighborhood $N(e)$ of instance $e$ is the set of instances of the same type as $e$ contained inside the neighborhood space $V_{N(e)}$. 

\begin{definition}(Density \cite{ref1284:Huang2008})
	For a given spatiotemporal space $V$, event type $f$ and its events instances in $D$, density is defined as follows:
	\begin{equation}
	\begin{split}
	Density(f, V) = \frac{|\{e | e.type = f \land \text{e is inside V}\}|}{|V|}
	\end{split}
	\end{equation}
	density is the number of instances of type $f$ occurring inside some space $V$ divided by the volume of space $V$.
	\label{Def:Density}
\end{definition}

\begin{definition}(Density ratio \cite{ref1284:Huang2008})
	Density ratio for two event types $f_{i_1}, f_{i_2}$ is defined as follows:
	\begin{equation}
	\begin{split}
	DensityRatio(f_{i_1} \rightarrow f_{i_2}) = \frac{avg_{(e.type = f_{i_1})}(Density(f_{i_2}, V_{N(e)}))}{Density(f_{i_2}, V)}
	\end{split}
	\end{equation} 
	$avg_{(e.type = f_{i_1})}(Density(f_{i_2}, V_{N(e)}))$ specifies the average density of instances of type $f_{i_2}$ occurring inside the neighborhood spaces $V_{N(e)}$ defined for instances, whose type is $f_{i_1}$. $V$ denotes the whole considered spatiotemporal space and $Density(f_{i_2}, V)$ specifies density of instances of type $f_{i_2}$ in space $V$.
	\label{Def:DensityRatio}
\end{definition}

Density ratio for the pattern $f_{i_1} \rightarrow f_{i_2}$ is defined as the ratio of the average density of instances of type $f_{i_2}$ inside the neighborhood spaces $V_{N(e)}$ defined for each instance $e \in f_{i_1}$ and the overall density of instances of type $f_{i_2}$ in the whole spatiotemporal space $V$.  

Density ratio expresses the strength of the attraction relation $f_{i_1} \rightarrow f_{i_2}$ between any two event types. If its value is greater than one, then occurrences of instances of type $f_{i_1}$ attract in their spatiotemporal neighborhood spaces occurrences of instances of type $f_{i_2}$. If its value is equal to one, then there is no correlation between these two event types. The value below one indicates negative relation (instances of type $f_{i_1}$ repel in their spatiotemporal neighborhood spaces occurrences of instances of type $f_{i_2}$). However, it may be very difficult to provide its appropriate value taking into account usually arbitrarily chosen $V$ and $V_{N(e)}$. Careful study of experimental results provided in \cite{ref1284:Huang2008} show some deficiencies in that subject.  

\begin{definition}(Sequence $\overrightarrow{s}$ and tailEventSet($\overrightarrow{s}$) \cite{ref1284:Huang2008})
	$\overrightarrow{s}$ denotes a k-length sequence of event types: $s[1] \rightarrow s[2] \rightarrow \dots \rightarrow s[k-1] \rightarrow s[k]$, where $ s[i] \in F $. 
	
	tailEventSet($\overrightarrow{s}$) is the set of instances of type $\overrightarrow{s}[k]$ participating in the sequence $\overrightarrow{s}$.
	\label{Def:Sequence}
\end{definition}

If a $k$-length sequence $\overrightarrow{s}$ is considered to be expanded, then the neighborhood spaces $V_{N(e)}$ and neighborhoods $N(e)$ are calculated for each $e \in$ tailEventSet($\overrightarrow{s}$).

\begin{definition}(Sequence index \cite{ref1284:Huang2008})
	For a given k-length sequence $\overrightarrow{s}$, sequence index is denoted as follows:
	\begin{enumerate}
		\item If $k = 2$ then:
		\begin{equation} 
		SeqIndex(\overrightarrow{s}) = DensityRatio(\overrightarrow{s}[1] \rightarrow \overrightarrow{s}[2])
		\end{equation}
		\item If $k > 2$ then:
		\begin{equation} 			
		SeqIndex(\overrightarrow{s}) = \text{min} \left \{
		\begin{array}{ll}
		SeqIndex(\overrightarrow{s}[1:k-1]),  \\
		DensityRatio(\overrightarrow{s}[k-1] \rightarrow \overrightarrow{s}[k])
		\end{array}
		\right.
		\end{equation}
	\end{enumerate}
	\label{Def:SequenceIndex}
\end{definition}

\begin{definition}
	Sequence (sequential pattern) $ \overrightarrow{s} $ is significant, if $ SeqIndex(\overrightarrow{s})$ $ \geq \theta $, where $ \theta $ is significance threshold given by the user.
	\label{Def:SignificantSeq}
\end{definition}

The value of $ SeqIndex(\overrightarrow{s}) $ for a $k$-length sequence $\overrightarrow{s}$ denotes the minimal strength of the $\rightarrow$ relation between any two consecutive event types participating in $\overrightarrow{s}$. In overall, sequence index provides means for discovering sequences for which its value is greater than the threshold given by the user. Possible significant sequences for the dataset shown in Fig.~\ref{Fig:1} are: $\overrightarrow{s_1} = (A \rightarrow B \rightarrow C \rightarrow D)$, $\overrightarrow{s_2} = (B \rightarrow C \rightarrow D)$ and $\overrightarrow{s_3} = (C \rightarrow D)$.

To illustrate the above presented notions and give an outline of the algorithm discovering sequential patterns, please consider the sequence $\overrightarrow{s_1} = A \rightarrow B \rightarrow C \rightarrow D$. Initially, the strength of the following relation $A \rightarrow B$ will be considered. The number of instances of type $B$ is significantly greater in the neighborhood spaces created for instances of type $A$ than the number of instances of type $B$ in the whole space $V$. So we assume significant value of density ratio for the relation $A \rightarrow B$ and create a sequence $\overrightarrow{s} = A \rightarrow B$. The tail event set of $\overrightarrow{s}$ will be $tailEventSet(\overrightarrow{s}) = \{b1, b2, \dots, b18\}$. The sequence $\overrightarrow{s}$ may be then expanded with event type $C$, because the average density of instances of type $C$ is significantly high in the neighborhood spaces defined for the instances of type $B$ contained in $tailEventSet(\overrightarrow{s})$. Similarly to the above, $\overrightarrow{s}$ will be expanded with event type $D$ to create $\overrightarrow{s_1}$.

Discovering all significant sequences is the aim of the algorithm ST-Miner \cite{ref1284:Huang2008}. The algorithm expands all significant sequences starting with 1-length sequences (singular event types). Each sequence is expanded in a dept-first manner by possible appending other event types. If the actual value of the sequence index is below a predefined threshold, then the sequence is not expanded any more. The crucial step in the expanding process is to compute $N(e)$ and density for each instance $e$ in the tail event set of the processed sequence. Computing $N(e)$ may be seen as performing spatial join and for this purpose we use the \textit{plane sweep algorithm} proposed in \cite{ref1284:Arge1998}. The complexity of the algorithm is $O(\mathcal{N} \sqrt{(\mathcal{N})})$ where $\mathcal{N}$ is the number of instances in $ D $.

\section{Discovering of Sequential Patterns by Means of Microclustering Approach} 
\label{Sec:MircoClust}

\begin{figure}[h!t]
	\centering
	\includegraphics[width=1.0\linewidth]{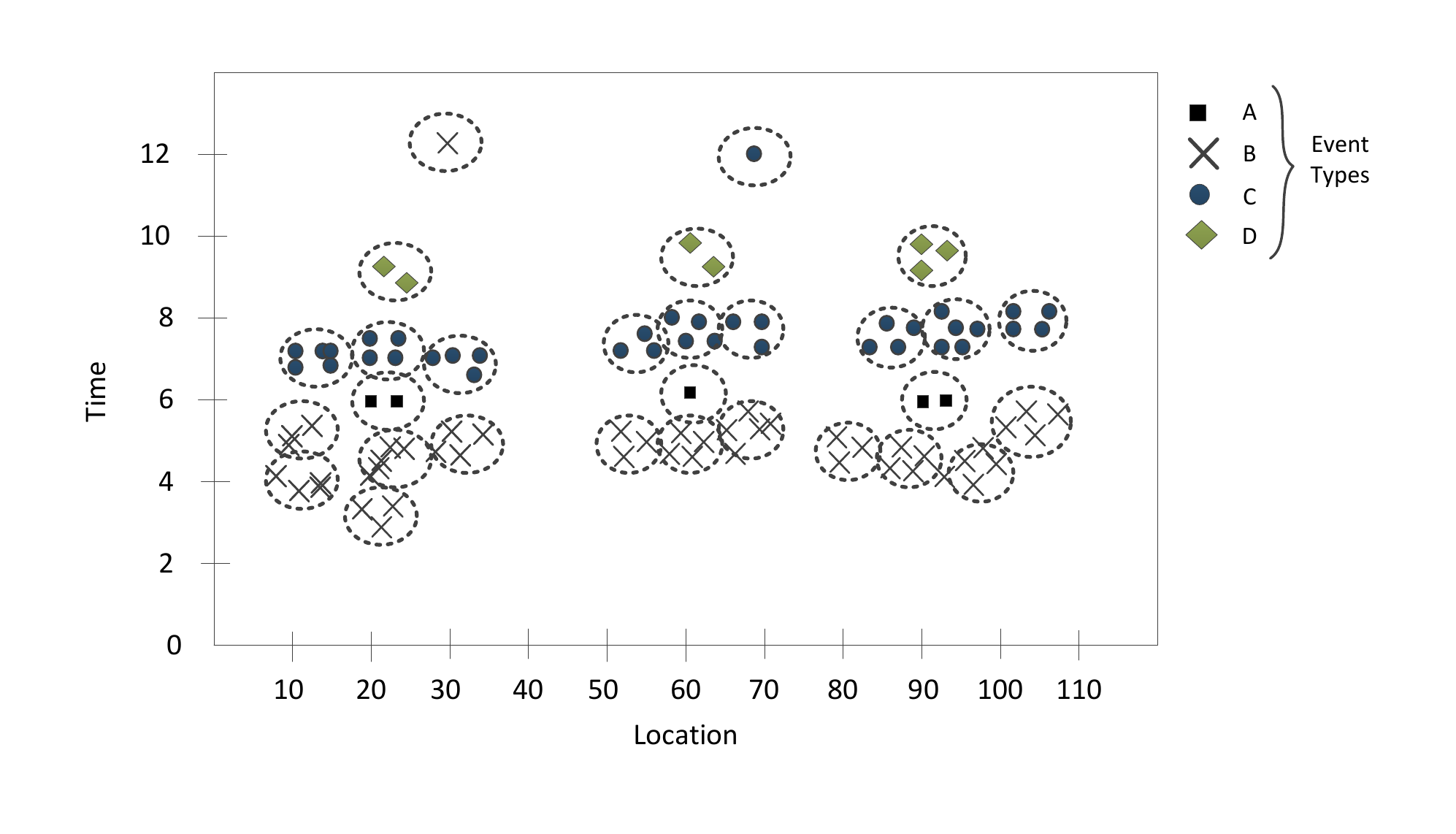}
	\caption[]{A microclustering approach applied to the problem of reducing the size of a dataset}
	\label{Fig:3}
\end{figure}

Searching for the neighborhoods $N(e)$ by means of the plane sweep algorithm is a costly operation. Initially, for each event type and its instances $e$ in $D$, a neighborhood $N(e)$ has to be computed. In the next steps, the number of computed neighborhoods will depend on the cardinalities of tail events sets of already discovered sequences. To resolve the mentioned problem we propose to use a \textit{microclustering} approach. The aim of the approach is to reduce the number of instances of different types by grouping neighboring instances of the same type into one microcluster. Sequential patterns discovery may then proceed using the modified dataset. In Fig.~\ref{Fig:3} we show a conception of microclustering.

Microcluster $ e_c $ is a quintuple consisting of \textit{microcluster identifier (CID), event type} of contained instances, \textit{a set of instances} contained in the microcluster, \textit{number of instances contained in the microcluster} and \textit{representative location} (for spatial and temporal) aspects. Representative location in both spatial and temporal dimensions is computed as a mean from respectively spatial coordinates and timestamps of instances contained in the microcluster.

Microclustering index $ MC_D $ created for dataset $ D $, is a structure containing set of created microclusters and it may be used for appropriate computation of density ratios and sequence indexes on the reduced dataset. According to the proposed method, tail event sets of sequences contain microclusters rather than the original events instances. 

Now we will propose our algorithm which will be based on the above introduced concept of microclustering. We need to appropriately modify definitions of density ratio and define facilitating indexing structures. We provide our modifications of the ST-Miner algorithm.

\begin{table}[h!t]
	\caption{An example of the microclustering index created by means of Algorithm~\ref{Alg:1} applied for the dataset shown in Fig.~\ref{Fig:1}}
	\begin{tabular}{c|c|c|c|c}
		\hline 
		CID & Event type & Contained instances & Num. of instances & Location (spatial, temporal) \\ 
		\hline \hline
		1 & A & $ \{a1\} $ & 1 & (15, 2) \\
		\hline 
		2 & A & $ \{a2\} $ & 1 & (51, 4) \\
		\hline
		\vdots & \vdots & \vdots & \vdots & \vdots \\
		\hline
		5 & B & $ \{b1, b2, b3, b4\} $ & 4 & (15, 3) \\
		\hline
		6 & B & $ \{b5, b6, b7, b8, b16\} $ & 5 & (51, 5) \\
		\hline 
		\vdots & \vdots & \vdots & \vdots & \vdots \\
		\hline
		10 & C & $ \{c1, c2, c3\} $ & 3 & (16, 6) \\
		\hline
		11 & C & $ \{c4, c5, c6\} $ & 3 & (55, 6) \\
		\hline
		\vdots & \vdots & \vdots & \vdots & \vdots \\
		\hline
		17 & D & $ \{d1, d2, d3\} $ & 3 & (15, 8) \\
		\hline
		18 & D & $ \{ d4 \} $ & 1 & (51, 8) \\
		\hline 
		\vdots & \vdots & \vdots & \vdots & \vdots \\
		\hline
	\end{tabular} 
	\label{Table:2}
\end{table} 

Let us consider the dataset given in Fig.~\ref{Fig:1}. The corresponding microclustering index is shown in Table~\ref{Table:2}. For the microclustering step one may apply similar approach as proposed in \cite{ref1284:Zhang1996}. The size of a microcluster may be limited by specifying its diameter threshold. In particular, for a given microcluster $e_c$ and its contained instances $\{e_1, e_2, \dots, e_n\}$, as the size of a microcluster we propose to use its diameter defined according to Formula (\ref{for:diameter}). By $\overrightarrow{e_i}$ we denote a vector containing location in spatial and temporal dimensions of a particular instance $e_i$. The proposed microclustering approach is presented in Algorithm \ref{Alg:1}. Algorithm \ref{Alg:1} has been inspired by the phase of building CF tree proposed in \cite{ref1284:Zhang1996}.

\begin{equation}
\begin{split}
&\mathcal{D}(e_c) = \sqrt{\frac{\sum_{i=1}^{m}\sum_{j=1}^{m}(distance(e_i - e_j))^2}{m*(m-1)}} \\
\end{split}
\label{for:diameter}
\end{equation}

where $distance(e_i - e_j)$ is the Euclidean distance between any two instances $e_i$ and $e_j$, $ i, j = 1 \dots m $ and $ m $ is the actual number of instances contained in $ e_c $. If needed, the spatial and temporal dimensions may be normalized for the microclustering step. In such a case, user may provide two diameter components $d_s$ for spatial dimension and $d_t$ for temporal dimension. The overall diameter threshold ($ d $) may be calculated as a normalized combination of these two values. 

\begin{algorithm}[h!t]
	\caption{Algorithm for microclustering dataset $D$ using Eq.~\ref{for:diameter}}	
	\begin{algorithmic}[1]
		\Require $ d $ - diameter threshold, $ D $ - dataset of instances.
		\Ensure $ MC_D $ - microclustering index for $ D $.
		\For {each event type $f \in F$}
		\State Take first instance  $ e $ from $D$.
		\State Initialize first microcluster by inserting instance $ e $.
		\While {There are instances in dataset $D$} 
		\State Take next $ e $ from dataset $D$ and insert it to the nearest $e_c$.
		\If {$\mathcal{D}(e_c) > d$ } 
		\State Split($e_c$) \Comment{Refer to Algorithm~\ref{Alg:1Split}.}
		\EndIf 
		\EndWhile
		\EndFor 
	\end{algorithmic} \label{Alg:1}
\end{algorithm}

\begin{algorithm}[h!t]
	\caption{Procedure for splitting microcluster Split($e_c$)}	
	\begin{algorithmic}[1]
		\Require $ e_c $ - microcluster to be split. 
		\Ensure Two microclusters $ e_c^{(i)} $ and $ e_c^{(j)} $ obtained after splitting $ e_c $.
		\State Find a pair of farthest instances in $e_c$.
		\State Mark the pair as seeds of two new microclusters $ e_c^{(i)} $ and $ e_c^{(j)} $.
		\State The rest of instances in $e_c$ assign to the nearest from $ e_c^{(i)} $ and $ e_c^{(j)} $.
	\end{algorithmic} \label{Alg:1Split}
\end{algorithm}

For each event type and for the first instance of particular event type initialize a microcluster. For each of next instances find the nearest microcluster and insert instance to that microcluster. If diameter of that microcluster is greater than given threshold, split microcluster according to Algorithm~\ref{Alg:1Split}. The split process proceeds as follows: find a pair of farthest instances in a microcluster considered to be split and mark the pair as seeds for two new microclusters. Redistribute the rest of instances from the original microcluster by assigning them to the nearest from two new microclusters.

\begin{figure}[h!t]
	\centering
	\includegraphics[width=1.0\linewidth]{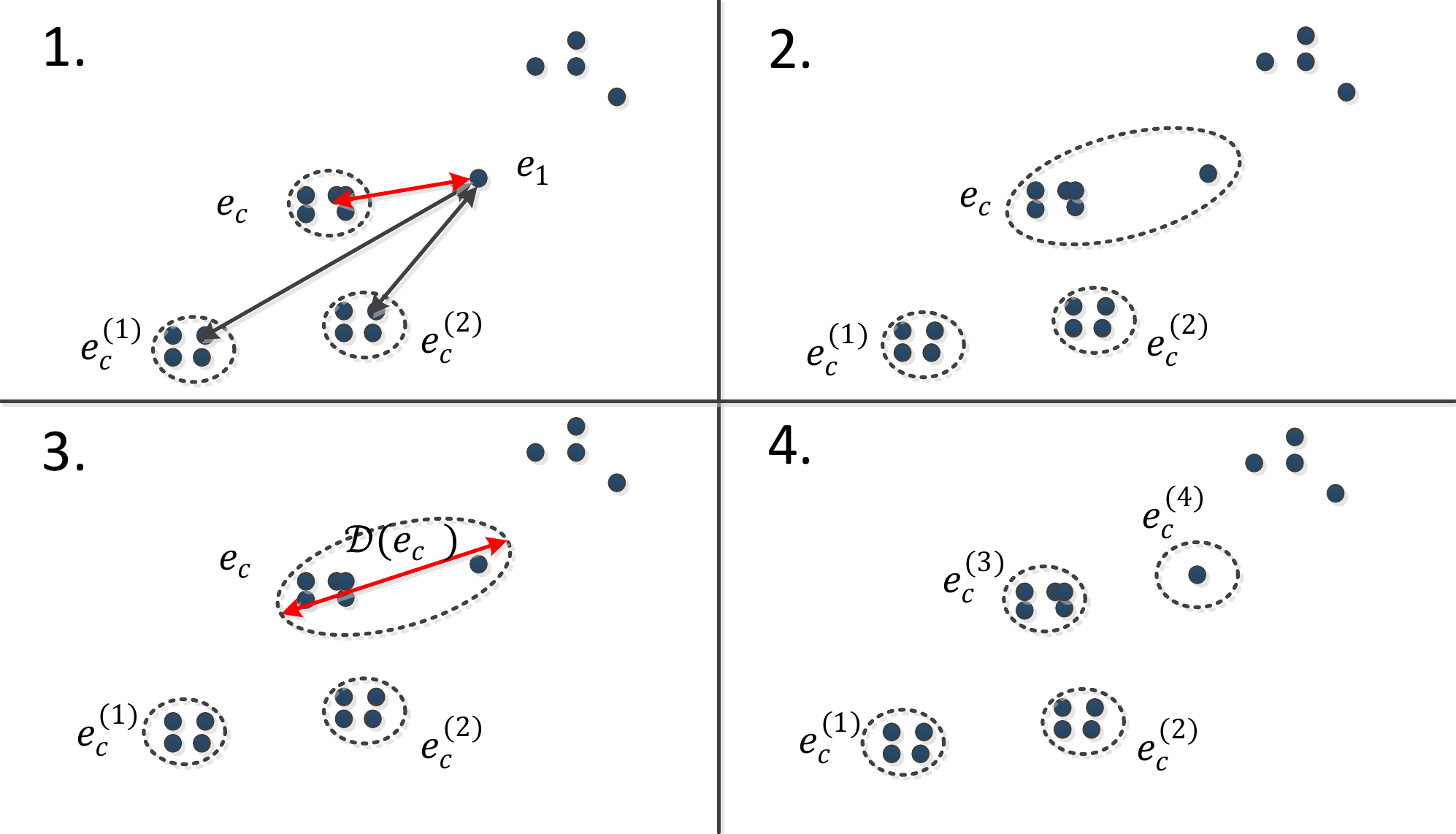}
	\caption[]{The four phases of Algorithm \ref{Alg:1}}
	\label{Fig:4}
\end{figure}

Algorithms \ref{Alg:1} and \ref{Alg:1Split} are illustrated in Fig.~\ref{Fig:4}. Definitions introduced in Section \ref{Sec:Basic notions} should be appropriately modified according to the above proposed concept of microclustering. By $MC_D$ we denote a microclustering index created for a dataset $D$. For any entry $e_c \in MC_D$ we propose two definitions of neighborhoods $n(e_c)$ and $N(e_c)$.

\begin{definition}
	For a microcluster $e_c' \in MC_D$, the neighborhood $n(e_c')$ of $e_c^{'}$ is defined as follows:
	\begin{equation}
	\begin{split}
	n(e_c') = & \{e_c | e_c \in MC_D \land distance(e_c'.location, e_c.location) \leq \mathcal{R} \\ 
	& \land (e_c'.time - e_c.time) \leq \mathcal{T} \land e_c'.type = e_c.type\}
	\end{split}
	\end{equation}
	that is, the neighborhood $ n(e_c') $ of a given microcluster $e_c'$ is defined as the set of those microclusters, whose centers are contained inside space $V_{N(e_c')}$ specified with respect to parameters $\mathcal{R}$ and $\mathcal{T}$. $ e_c'.location $ and $e_c.location$ refer respective to representative spatial locations of $ e_c' $ and $ e_c $.
	\label{Def:n}
\end{definition}

\begin{definition}
	For a microcluster $e_c'$ and for its neighborhood $n(e_c')$, the set of instances covered by entries from $n(e_c')$ is defined as follows:
	\begin{equation}
	\begin{split}
	N(e_c') = & \{p | p \in D \land p \text{ is inside } e_c \land e_c \in n(e_c') \land e_c'.type = e_c.type \}
	\end{split}
	\end{equation}
	that is, as the neighborhood $ N(e_c') $ of a given microcluster $ e_c' $ we define a set of those instances from $D$, that each of which is contained in the microclusters from $n(e_c')$. 
	\label{Def:N}
\end{definition}

The above introduced definitions $n(e_c)$ and $N(e_c)$ are useful in reformulating definitions of density and density ratio introduced in Section \ref{Sec:Basic notions}. By $MC_D(f)$ we define a set of entries in $MC_D$ of event type $f$. 

In Definitions \ref{Def:n} and \ref{Def:N} we do not refer to any particular event type, which microclusters or instances occur respectively in $n(e_c)$ or $N(e_c)$. However, it may be noticed from already provided notions, that density ratio considered for any two event types always take into account only neighborhoods of those particular event types. For example, considering the pattern $f_{i_1} \rightarrow f_{i_2}$, for any $e_c \in MC_D(f_{i_1})$, $n(e_c)$ and $N(e_c)$ contain respectively only microclusters and instances of type $f_{i_2}$.  

\begin{remark}
	For a microcluster $e_c \in MC_D$, the number of instances contained in this microcluster is denoted by $|e_c|$. For example, the number of instances contained in microcluster $e_{c}^{(5)}$ from Table \ref{Table:2} is $|e_{c}^{(5)}| = 4$. 
\end{remark}

\begin{definition}{Modified density.}
	For a given spatiotemporal space $V$, event type $f$ and its set of microclusters $MC_D(f) = \{e_c^{(1)}, e_c^{(2)}, \dots, e_c^{(n)}\}$ contained inside $V$, the modified density is defined as follows:
	\begin{equation}
	\begin{split}
	density_{MC_D}(f, V) = & \frac{|e_c^{(1)}| + |e_c^{(2)}| + \dots + |e_c^{(n)}|}{|V|}, \\
	& \text{ where } \{e_c^{(1)}, e_c^{(2)}, \dots, e_c^{(n)}\} \text{ inside } V 
	\end{split}
	\end{equation}
	that is, we define modified density as the quotient of the number of instances contained in $e_c^{(i)} \in MC_D(f)$ and the volume of space $V$. We say that a microcluster $ e_c $ is inside space $V$, if its representative location is inside $V$.
\end{definition}

\begin{definition}{Modified density ratio.}
	For two event types $ f_{i_1}, f_{i_2} $ and their entries in microclustering index $MC_D$, the modified density ratio is defined as follows:
	\begin{equation}
	\begin{split}
	DensityRatio_{MC_D}(f_{i_1} \rightarrow f_{i_2}) = \frac{\sum_{e_c.type= f_{i_1}}\big(|e_c|*density_{MC_D}(f_{i_2}, V_{N(e_c)})\big)}{\big(\sum_{e_c.type =  f_{i_1}}|e_c|\big)*density_{MC_D}(f_{i_2}, V)}
	\end{split}
	\end{equation}
	comparing to Definition \ref{Def:DensityRatio}, the above definition takes into account proposed microclustering index and its entries.
	\label{Def:ModDensityRatio}
\end{definition}

The idea of Definition \ref{Def:ModDensityRatio} is to calculate density ratio for the relation $f_{i_1} \rightarrow f_{i_2}$ using entries in microclustering index $MC_D$. Taking a microcluster $e_c \in MC_D(f_{i_1})$ and its neighborhood space $V_{n(e_c)}$, $density_{MC_D}(f_{i_2}, V_{N(e_c)})$ is weighted by the number $|e_c|$, that is by the number of instances contained in $e_c$. That is, the term: 

$$\frac{\sum_{e_c.type = f_{i_1}}\big(|e_c|*density_{MC_D}(f_{i_2}, V_{N(e_c)})\big)}{\big(\sum_{e_c.type = f_{i_1}}|e_c|\big)}$$

specifies the weighted average density of instances of type $f_{i_2}$ occurring in the neighborhood spaces $V_{N(e_c)}$ defined for microclusters $e_c \in MC_D(f_{i_1})$ with weights $|e_c|$. Before we proceed with an algorithm description we need to introduce one more definition.

\begin{definition}
	For a sequence $\overrightarrow{s} \rightarrow f$ of length $k + 1$, that $f$ follows event type $\overrightarrow{s}[k]$. tailEventSet($ \overrightarrow{s} \rightarrow f $) is equal to the the set of microclusters of type $f$ in $MC_D$ following the sequence $\overrightarrow{s}$. This fact is denoted by the formula: tailEventSet($ \overrightarrow{s} \rightarrow f $) $ = $ $  \{n(e_c^{(1)}) \cup n(e_c^{(2)}) \cup \dots \cup n(e_c^{(n)}) \} $ where tailEventSet($ \overrightarrow{s} $) = $ \{e_c^{(1)}, e_c^{(2)}, \dots, e_c^{(n)}\} $.
	\label{Def:ModSequenceAndTail}
\end{definition}

\begin{figure}[h!t]
	\centering
	\includegraphics[width=1.0\linewidth]{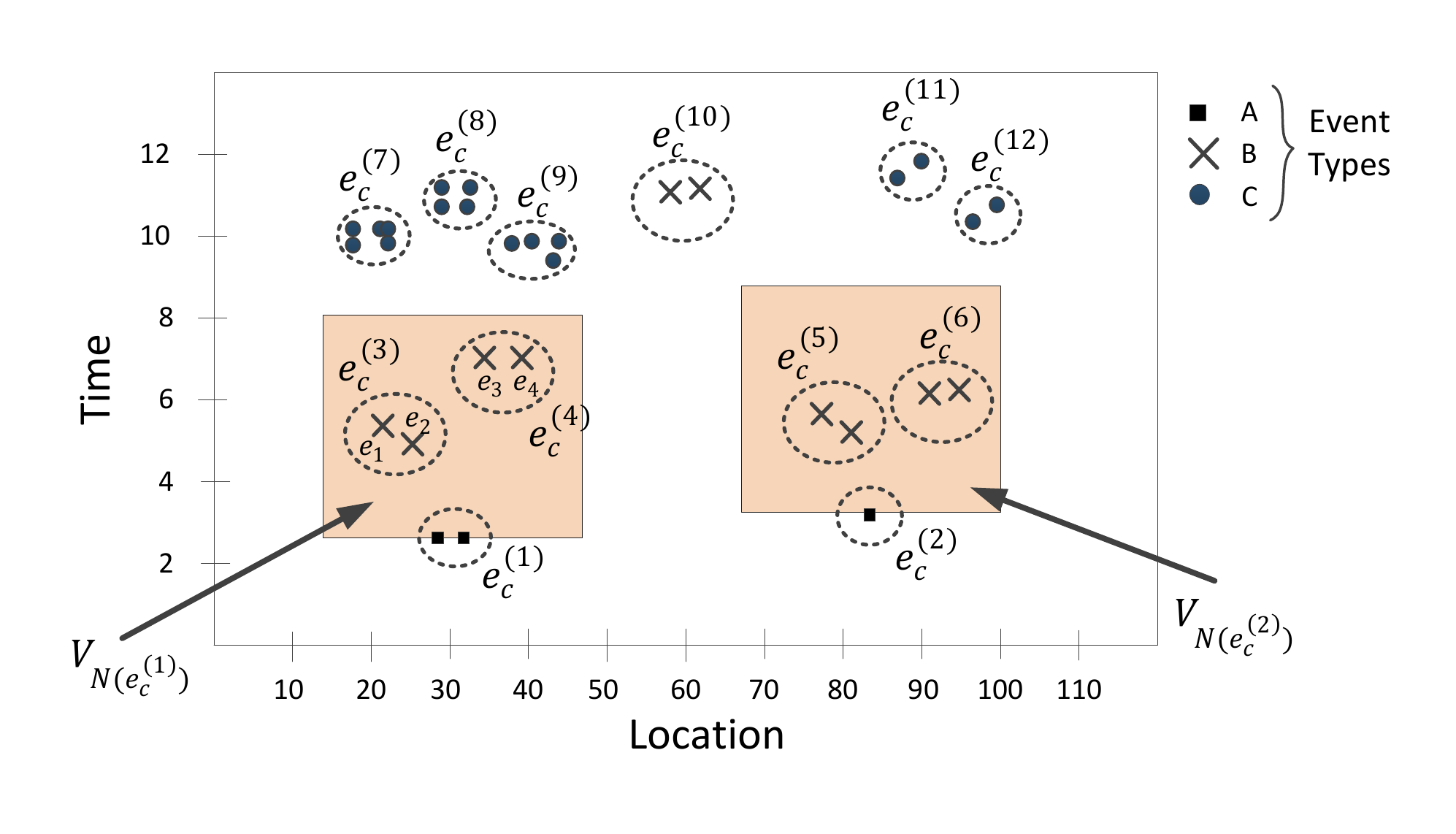}
	\caption[]{An example of the dataset containing microclusters and their neighborhoods}
	\label{Fig:5}
\end{figure}

To illustrate above notions, please consider Fig.~\ref{Fig:5} and assume that we investigate the relation $A \rightarrow B$. Assume that at the beginning $\overrightarrow{s} = A$ and $\overrightarrow{s}$ should be expanded with $B$, that is consider $\overrightarrow{s} \rightarrow B$. $tailEventSet(\overrightarrow{s}) = \{e_{c}^{(1)}, e_{c}^{(2)}\}$ and $tailEventSet(\overrightarrow{s} \rightarrow B)  $$ = n(e_c^{(1)}) \cup n(e_c^{(2)}) $$ = \{e_{c}^{(3)}, e_{c}^{(4)}, e_{c}^{(5)}, e_{c}^{(6)}\}$. Additionally: $n(e_c^{(1)} = \{e_{c}^{(3)}, e_{c}^{(4)}\})$ and $N(e_c^{(1)}) = \{e_1, e_2, e_3, e_4\}$.

Now we proceed with the description of the proposed Micro-ST-Miner algorithm. For a given dataset $D$ containing events instances and their types, Algorithm \ref{Alg:2} starts with computing microclustering index $MC_D$. Then, for any event type $f$, the algorithm creates 1-length sequence $\overrightarrow{s}$ and expands it using the \textit{ExpandSequence} procedure (our modification of the procedure already proposed in \cite{ref1284:Huang2008}). 

\begin{algorithm}[h!t]
	\caption{Micro-ST-Miner - algorithm for discovering sequential patterns from event-based spatiotemporal data using microclustering index}	
	\begin{algorithmic}[1]
		\Require $D$ - a dataset containing events instances, $ \theta $ - significance threshold for sequence index.
		\Ensure A set of significant sequential patterns.
		\State Perform microclustering on the dataset $ D $ resulting in a microclustering index $ MC_{D} $.
		\For {each event type $ f $ and its entries in $ MC_{D}(f) $}    
		\State Create 1-length sequence $ \overrightarrow{s} $ from $ f $.
		\State tailEventSet($ \overrightarrow{s} $) $ := $ $ MC_D(f) $.
		\State \textit{ExpandSequence($ \overrightarrow{s} $, $ MC_D $,  $ \theta $)}.
		\EndFor 
	\end{algorithmic} \label{Alg:2}
\end{algorithm}

\begin{algorithm}[h!t]
	\caption{ExpandSequence($ \protect \overrightarrow{s} $, $ MC_D $, $ \theta $) procedure}
	\begin{algorithmic}[1]
		\Require $\overrightarrow{s}$ - a sequence of event types, $ MC_D $ - microclustering index created for $D$, $\theta$ - threshold for sequence index.
		\For {each event type $ f $ and its entries in $ MC_{D} $}
		\State Calculate SequenceIndex($ \overrightarrow{s} \rightarrow f $).
		\If {value of the sequence index $ \geq \theta$ }
		\State Remember $ \overrightarrow{s} \rightarrow f $ as significant.
		\State \begin{varwidth}[t]{\linewidth} 
			tailEventSet($ \overrightarrow{s} \rightarrow f$) $ := $ $\{n(e_c^{(1)}) \cup n(e_c^{(2)}) \cup \dots \cup n(e_c^{(n)}) \}$ \par 
			\hskip\algorithmicindent where tailEventSet($ \overrightarrow{s} $) = $ \{e_c^{(1)}, e_c^{(2)}, \dots, e_c^{(n)}\} $.
		\end{varwidth}
		\State ExpandSequence($ (\overrightarrow{s} \rightarrow f) $, $ MC_D $, $ \theta $).
		\EndIf 
		\EndFor  
	\end{algorithmic}
	\label{Alg:3}
\end{algorithm}

\begin{algorithm}[h!t]
	\caption{Calculate SequenceIndex($\protect \overrightarrow{s} \rightarrow f$) procedure}
	\begin{algorithmic}[1]
		\Require $ \overrightarrow{s} \rightarrow f$ - a sequence of event types; $k + 1$ - the length of the sequence $\overrightarrow{s}  \rightarrow f$
		\State return min(SequenceIndex($\overrightarrow{s}$), DensityRatio($ \overrightarrow{s}[k] \rightarrow f $)).
	\end{algorithmic}
	\label{Alg:4}
\end{algorithm}

In the next sections, we consider the problem of more efficient performing microclustering step presented in Algorithm~\ref{Alg:1}. In particular, we propose to perform microclustering by means of triangle inequality, limit a number of instances contained in a microcluster or discuss other approaches to perform microclustering.

\subsection{Other Extensions to the Microclustering Approach}

In experimental results, we verify the usefulness of the approach where the size of each microgroup is limited by the number of contained instances. Algorithm~\ref{Alg:5} presents the modification, assuming that the maximal number of instances contained in a microcluster is denoted by $ K $.

\begin{algorithm}[h!t]
	\caption{Algorithm for microclustering dataset $D$ using Eq.~\ref{for:diameter}}	
	\begin{algorithmic}[1]
		\Require $ d $ - diameter threshold, $ D $ - dataset of instances, $ K $ - maximal number of instances in a microcluster.
		\Ensure $ MC_D $ - microclustering index for $ D $.
		\For {each event type $f \in F$}
		\State Take first instance  $ e $ from $D$.
		\State Initialize first microcluster by inserting instance $ e $.
		\While {There are instances in dataset $D$}
		\State Take next $ e $ from dataset $D$ and insert it to the nearest $e_c$.	
		\If {$ |e_c| > K $}
		\State Split($e_c$). \Comment{Refer to Algorithm~\ref{Alg:1Split}.}
			\If { $\mathcal{D}(e_c^{(i)}) > d$ } 
			\State Split($e_c^{(i)}$).
			\EndIf 
			\If {$\mathcal{D}(e_c^{(j)}) > d$}
			\State Split($e_c^{(j)}$).
			\EndIf 
		\ElsIf { $\mathcal{D}(e_c^{(j)}) > d$ } 
		\State Split($e_c$).
		\EndIf 
		\EndWhile
		\EndFor 
	\end{algorithmic} \label{Alg:5}
\end{algorithm}

This approach is motivated by the fact, that some datasets may contain instances with very similar values of spatial coordinates. Consider for example a dataset where occurrences of event types are fixed (in spatial aspect) on some coordinates of a grid. This is often case for datasets containing weather or climate data. In such a case, despite small value of diameter threshold, it may be possible that a microcluster contains a significant number of instances. Due to quadratic complexity of Eq.~\ref{for:diameter} such situation may be undesirable when calculating diameter of the microcluster. The aim of Algorithm~\ref{Alg:5} is to tackle such situation.

Algorithm~\ref{Alg:5} proceeds similarly to Algorithm~\ref{Alg:1}. Split operation is the same as shown in Algorithm~\ref{Alg:1Split}. When an instance is inserted to the nearest microcluster, Algorithm~\ref{Alg:1} checks if the number of instances in the microcluster is greater than the user given parameter $ K $. If this is true, the microcluster is split into two new microclusters. Diameters of these two microclusters are verified and, if this is needed, microclusters are split.

In future studies, it may be appropriate to adapt triangle inequality to more efficiently perform microclustering process \cite{ref1284:Kryszkiewicz2010,ref1284:Kryszkiewicz2010neighborhood}.

\section{Experimental Results}
\label{Sec:ExperimentalResults}

In Table~\ref{Table:3}, we explain the main parameters of data generator. Our data generator is similar to proposed in \cite{ref1284:Huang2008,ref1284:Agrawal1994}. Data generation method is as follows. For a spatio-temporal space $ V $ defined by parameters $ DSize $ and $ TSize $, the actual number of maximal patterns in a dataset is given by $ Pn$, the maximal length of generated patterns is given by $ Ps$. First, for each pattern we generate event types participating in this pattern. Event types are uniformly selected from the set of size $ Nf $. 

\begin{table}
	\caption{Description of data generator parameters}
	\begin{tabularx}{\textwidth}{cX}
		\toprule
		Parameter name & Description \\
		\midrule \midrule
		$Ps$ & Length of a maximal generated sequence \\
		$Pn$ & Number of maximal sequences in generated data \\
		$DSize$ & Size of spatial dimension of embedding space $ V $ \\
		$TSize$ & Size of temporal dimension of embedding space $ V $ \\
		$Nf$ & Total number of event types available in data generator \\
		$Ni$ & Starting number of instances of the first event type participating in a pattern  \\
		$\mathcal{R}$ & Size of spatial dimension of neighborhood space $ V_{N(e)} $ \\
		$\mathcal{T} $ & Size of temporal dimension of neighborhood space $ V_{N(e)} $ \\
		\bottomrule
	\end{tabularx}
	\label{Table:3}
\end{table} 

Patterns are generated as follows. For each event type in a pattern we generate $ Ni $ instances. Instances of the first event type are uniformly distributed in spatio-temporal space $ V $ of size $ DSize*DSize*Tsize $. For each instance of the next event type, we randomly select instance of the previous event type and put it into spatio-temporal neighborhood space of that instance. This means, that the number of instances participating in a pattern is given by $ Ps*Ni $. Additionally, we generate the same number of noise instances and distribute them uniformly in the embedding spatio-temporal space $ V $. In the result, each dataset contains total number of instances $ Ti = Ni * Ps * Pn * 2 $.

Please note, that the word \textit{maximal} means that generated pattern (sequence) contains also some subsequences that have to be discovered in a given dataset.

\subsection{Microclustering Step}

Algorithm~\ref{Alg:1} has been applied to different datasets. We perform microclustering for datasets varying with number of event sequences, patterns length and patterns number. In our experiments, we generated datasets similar to \cite{ref1284:Huang2008}. In Table~\ref{Table:4}, we show microclustering index size and compression ratio for datasets with different number of event sequences. Compression ratio is calculated as the ratio of average dataset size divided by the microclustering index size. In Table~\ref{Table:5}, we show how compression ratio and microclustering index size is dependent on the number of patterns in a dataset. In Table~\ref{Table:6}, we provide microclustering index size and compression ratio for datasets with varying maximal pattern length. 

\begin{figure}[h!t]
	\centering
	\includegraphics[width=0.75\linewidth]{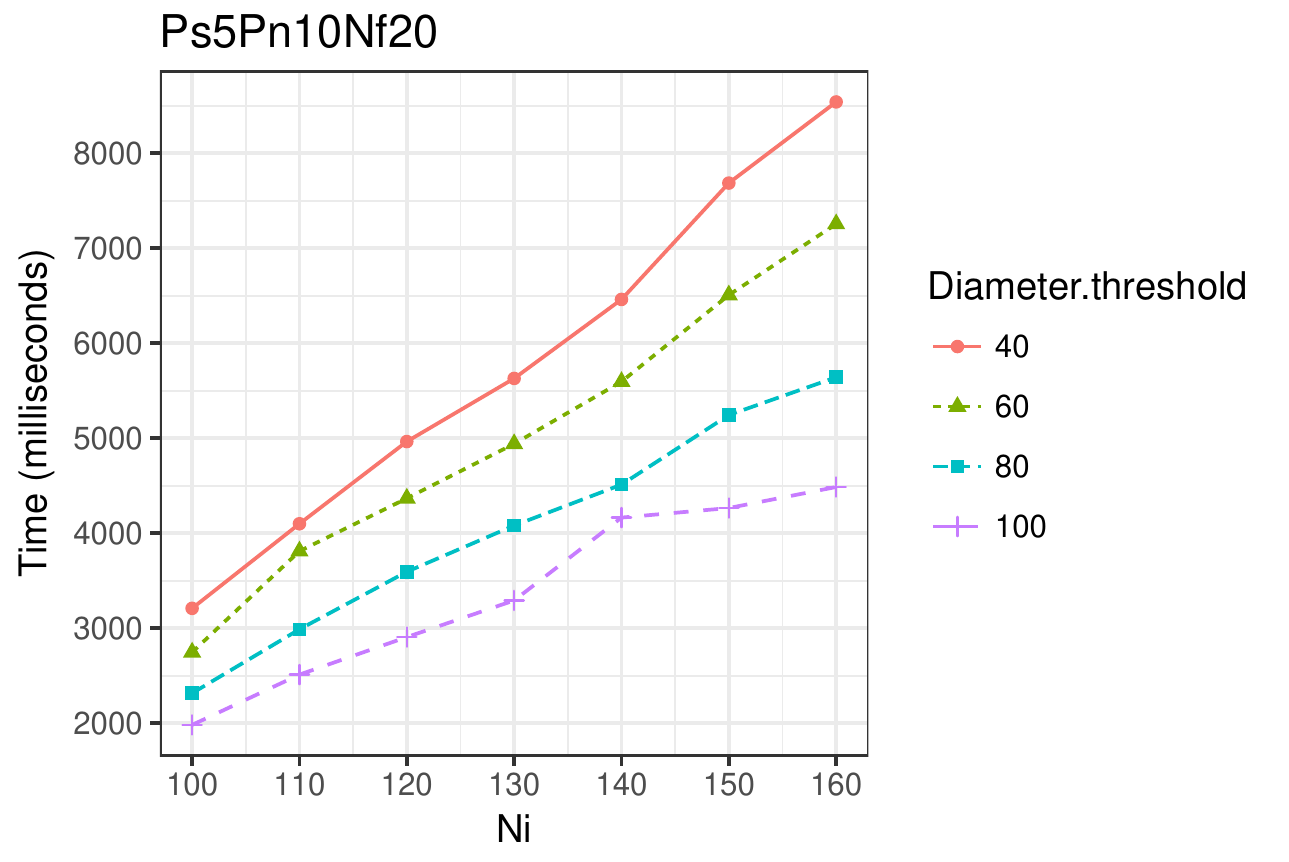}
	\caption[]{Microclustering time for dataset varying in number of event sequences}
	\label{Fig:6}
\end{figure}

\begin{figure}[h!t]
	\centering
	\includegraphics[width=0.75\linewidth]{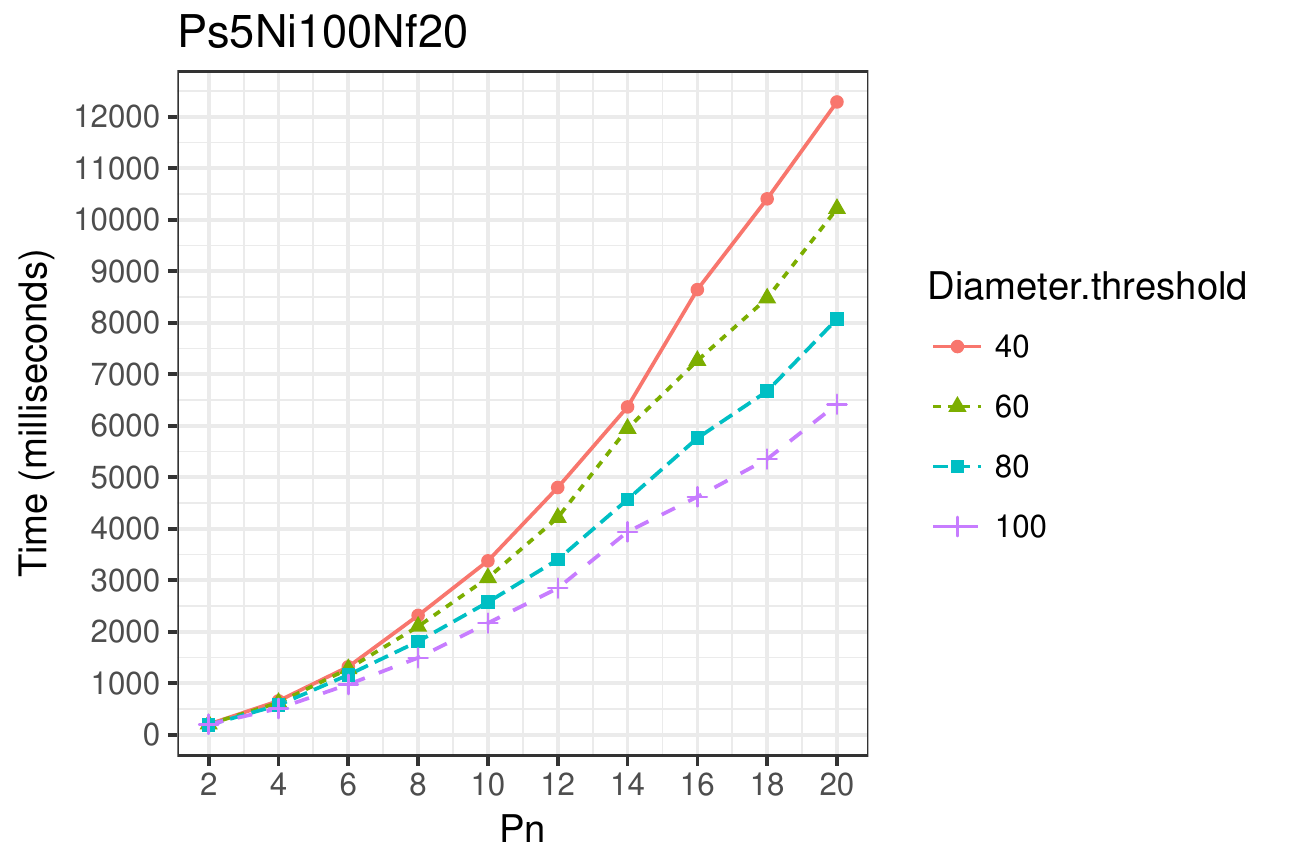}
	\caption[]{Microclustering time for dataset varying in patterns number}
	\label{Fig:7}
\end{figure}

\begin{figure}[h!t]
	\centering
	\includegraphics[width=0.75\linewidth]{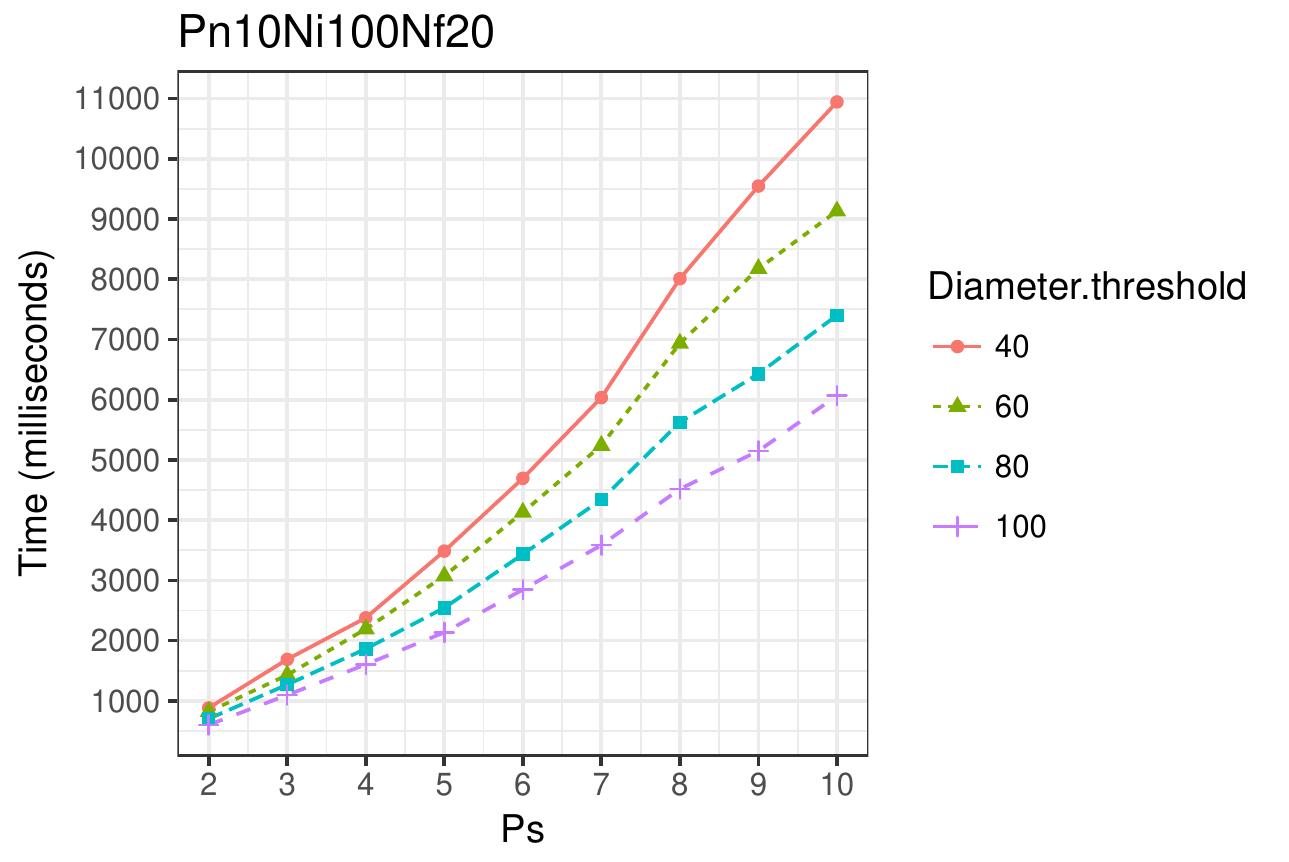}
	\caption[]{Microclustering time for dataset varying in patterns length}
	\label{Fig:8}
\end{figure}

Figures~\ref{Fig:6}, \ref{Fig:7} and \ref{Fig:8} contain microclustering time (in milliseconds) for these three types of datasets. We perform experiments using four diameter threshold values: $ 40, 60, 80, 100 $.  Results presented in Tables~\ref{Table:4}, \ref{Table:5}, \ref{Table:6} and Figures~\ref{Fig:6}, \ref{Fig:7}, \ref{Fig:8} are obtained using Algorithm~\ref{Alg:1}. We apply Algorithm~\ref{Alg:5} to real dataset.

\begin{table}[h!t]
	{\scriptsize 
		\caption{Average dataset size, averages of reduced dataset size and Compression Ratio after applying microclustering algorithm to datasets with different event sequences values}
		\begin{tabularx}{\textwidth}{ccc@{\hskip 7pt}XXXX}
			\toprule
			\multicolumn{7}{c}{Ps = 5, Pn = 10, Nf = 20, $\mathcal{R} = 10$, $\mathcal{T} = 10$, $DSize = 1000$, $TSize = 1200$ } \\
			\midrule \midrule
			Ni & Dataset size &  & \multicolumn{4}{c}{Microcluster's diameter threshold} \\
			\midrule \midrule 
			& & & 40 & 60 & 80 & 100  \\
			\cmidrule{4-7} 
			\addlinespace 
			\multirow{2}{*}{100} & \multirow{2}{*}{10000} & Microclustering index size & 7246 & 6502 & 5489 & 4451  \\
			& & Compression Ratio & 1.38007 & 1.53799 & 1.82183 & 2.24669	\\
			\midrule \addlinespace 
			\multirow{2}{*}{110} & \multirow{2}{*}{11000}    & Microclustering index size & 7916 & 7045 & 5862 & 4695 \\
			& & Compression Ratio  & 1.38959 & 1.56139 & 1.87649 & 2.34292\\
			\midrule \addlinespace 
			\multirow{2}{*}{120} &\multirow{2}{*}{12000}   & Microclustering index size & 8583 & 7564 & 6243 & 4947 \\
			& & Compression Ratio  & 1.39811 & 1.58646 & 1.92215 & 2.42571 \\
			\midrule \addlinespace 
			\multirow{2}{*}{130} &\multirow{2}{*}{13000}   & Microclustering index size & 9321 & 8129 & 6623 & 5184 \\
			& & Compression Ratio  & 1.3947 & 1.59921 & 1.96286 & 2.50772 \\
			\midrule \addlinespace 
			\multirow{2}{*}{140} &\multirow{2}{*}{14000}   & Microclustering index size & 9959 & 8619 & 6942 & 5399	 \\
			& & Compression Ratio  & 1.40576 & 1.62432 & 2.01671 & 2.59307	 \\
			\midrule \addlinespace 
			\multirow{2}{*}{150} &\multirow{2}{*}{15000}   & Microclustering index size & 10577 & 9101 & 7283 & 5601 \\
			& & Compression Ratio  & 1.41817 & 1.64817 & 2.05959 & 2.67809 \\
			\midrule \addlinespace 
			\multirow{2}{*}{160} & \multirow{2}{*}{16000}   & Microclustering index size & 11214 & 9560 & 7573 & 5787 \\
			& & Compression Ratio  & 1.4267 & 1.67364 & 2.11277 & 2.76482 \\
			\bottomrule
		\end{tabularx}
		\label{Table:4}	
	}
\end{table} 
	
\begin{table}[h!t]
	{\scriptsize 
	\caption{Average dataset size, averages of reduced dataset size and Compression Ratio after applying microclustering algorithm to datasets with different patterns number}
	\begin{tabularx}{\textwidth}{ccc@{\hskip 7pt}XXXX}
		\toprule
		\multicolumn{7}{c}{Ps = 5, Ni = 100, Nf = 20, $\mathcal{R} = 10$, $\mathcal{T} = 10$, $DSize = 1000$, $TSize = 1200$ } \\
		\midrule \midrule
		Pn & Dataset size &  & \multicolumn{4}{c}{Microcluster's diameter threshold} \\
		\midrule \midrule 
		& & & 40 & 60 & 80 & 100  \\
		\cmidrule{4-7} 
		\addlinespace 
		\multirow{2}{*}{2} & \multirow{2}{*}{2000} & Microclustering index size & 1512 & 1469 & 1396 & 1303    \\
		& & Compression Ratio & 1.32275 & 1.36147 & 1.43266 & 1.53492\\
		\midrule \addlinespace 
		\multirow{2}{*}{4} & \multirow{2}{*}{4000}    & Microclustering index size & 2977 & 2837 & 2599 & 2311 \\
		& & Compression Ratio  & 1.34363 & 1.40994 & 1.53905 & 1.73085\\
		\midrule \addlinespace 
		\multirow{2}{*}{6} &\multirow{2}{*}{6000}   & Microclustering index size & 4376 & 4085 & 3631 & 3126 \\
		& & Compression Ratio  & 1.37112 & 1.46879 & 1.65244 & 1.91939 \\
		\midrule \addlinespace 
		\multirow{2}{*}{8} &\multirow{2}{*}{8000}   & Microclustering index size & 5852 & 5349 & 4638 & 3855  \\
		& & Compression Ratio  & 1.36705 & 1.49561 & 1.72488 & 2.07523\\
		\midrule \addlinespace 
		\multirow{2}{*}{10} &\multirow{2}{*}{10000}   & Microclustering index size & 7209 & 6461 & 5452 & 4421 \\
		& & Compression Ratio  & 1.38715 & 1.54775 & 1.83419 & 2.26193\\
		\midrule \addlinespace 
		\multirow{2}{*}{12} &\multirow{2}{*}{12000}   & Microclustering index size & 8580 & 7562 & 6238 & 4933 \\
		& & Compression Ratio  & 1.3986 & 1.58688 & 1.92369 & 2.4326 \\
		\midrule \addlinespace 
		\multirow{2}{*}{14} & \multirow{2}{*}{14000}   & Microclustering index size & 9937 & 8611 & 6944 & 5379 \\
		& & Compression Ratio  & 1.40888 & 1.62583 & 2.01613 & 2.60271 \\
		\midrule \addlinespace 
		\multirow{2}{*}{16} & \multirow{2}{*}{16000}   & Microclustering index size & 11260 & 9589 & 7574 & 5785 \\
		& & Compression Ratio  & 1.4209 & 1.66858 & 2.11249 & 2.76577\\
		\midrule \addlinespace 
		\multirow{2}{*}{18} & \multirow{2}{*}{18000}   & Microclustering index size & 12516 & 10506 & 8180 & 6147 \\
		& & Compression Ratio  & 1.43816 & 1.71331 & 2.20049 & 2.92826 \\
		\midrule \addlinespace 
		\multirow{2}{*}{20} & \multirow{2}{*}{20000}   & Microclustering index size & 13867 & 11470 & 8766 & 6508 \\
		& & Compression Ratio  & 1.44221 & 1.74368 & 2.28154 & 3.07314\\
		\bottomrule
	\end{tabularx}
}
	\label{Table:5}	
\end{table}

\begin{table}[h!t]
	{\scriptsize 
		\caption{Average dataset size, averages of reduced dataset size and Compression Ratio after applying microclustering algorithm to datasets with different patterns length}
		\begin{tabularx}{\textwidth}{ccc@{\hskip 7pt}XXXX}
			\toprule
			\multicolumn{7}{c}{Pn = 10, Ni = 100, Nf = 20, $\mathcal{R} = 10$, $\mathcal{T} = 10$, $DSize = 1000$, $TSize = 1200$ } \\
			\midrule \midrule
			Ps & Dataset size &  & \multicolumn{4}{c}{Microcluster's diameter threshold} \\
			\midrule \midrule 
			& & & 40 & 60 & 80 & 100  \\
			\cmidrule{4-7} 
			\addlinespace 
			\multirow{2}{*}{2} & \multirow{2}{*}{4000} & Microclustering index size & 3505 & 3305 & 2998 & 2630 \\
			& & Compression Ratio & 1.14123 & 1.21029 & 1.33422 & 1.52091	 \\
			\midrule \addlinespace 
			\multirow{2}{*}{3} & \multirow{2}{*}{6000}    & Microclustering index size &  4883 & 4527 & 4000 & 3403 \\
			& & Compression Ratio  & 1.22875 & 1.32538 & 1.5 & 1.76315 \\
			\midrule \addlinespace 
			\multirow{2}{*}{4} &\multirow{2}{*}{8000}   & Microclustering index size & 6142 & 5600 & 4838 & 3986 \\
			& & Compression Ratio  & 1.30251 & 1.42857 & 1.65358 & 2.00702 \\
			\midrule \addlinespace 
			\multirow{2}{*}{5} &\multirow{2}{*}{10000}   & Microclustering index size & 7227 & 6505 & 5501 & 4451 \\
			& & Compression Ratio  &  1.3837 & 1.53728  & 1.81785 & 2.24669 \\
			\midrule \addlinespace 
			\multirow{2}{*}{6} &\multirow{2}{*}{12000}   & Microclustering index size & 8302 & 7327 & 6068 & 4822 \\
			& & Compression Ratio  & 1.44543 & 1.63778 & 1.97759 & 2.48859	\\
			\midrule \addlinespace 
			\multirow{2}{*}{7} &\multirow{2}{*}{14000}   & Microclustering index size & 9320 & 8099 & 6597 & 5141 \\
			& & Compression Ratio  & 1.50215 & 1.72861 & 2.12218 & 2.72321 \\
			\midrule \addlinespace 
			\multirow{2}{*}{8} & \multirow{2}{*}{16000}   & Microclustering index size & 10372 & 8903 & 7126 & 5484 \\
			& & Compression Ratio  & 1.54261 & 1.79715 & 2.2453 & 2.91758	 \\
			\midrule \addlinespace 
			\multirow{2}{*}{9} & \multirow{2}{*}{18000}   & Microclustering index size &  11373 & 9630 & 7586 & 5755 \\
			& & Compression Ratio  & 1.5827 & 1.86916 & 2.37279 & 3.12772 \\
			\midrule \addlinespace 
			\multirow{2}{*}{10} & \multirow{2}{*}{20000}   & Microclustering index size & 12258 & 10272 & 7993 & 6003\\
			& & Compression Ratio  & 1.63159 & 1.94704 & 2.50219 & 3.33167	\\
			\bottomrule
		\end{tabularx}
	}
	\label{Table:6}	
\end{table} 

\subsection{Results for the Micro-ST-Miner Algorithm}

\begin{figure}[h!t]
	\centering
	\includegraphics[width=0.98\linewidth]{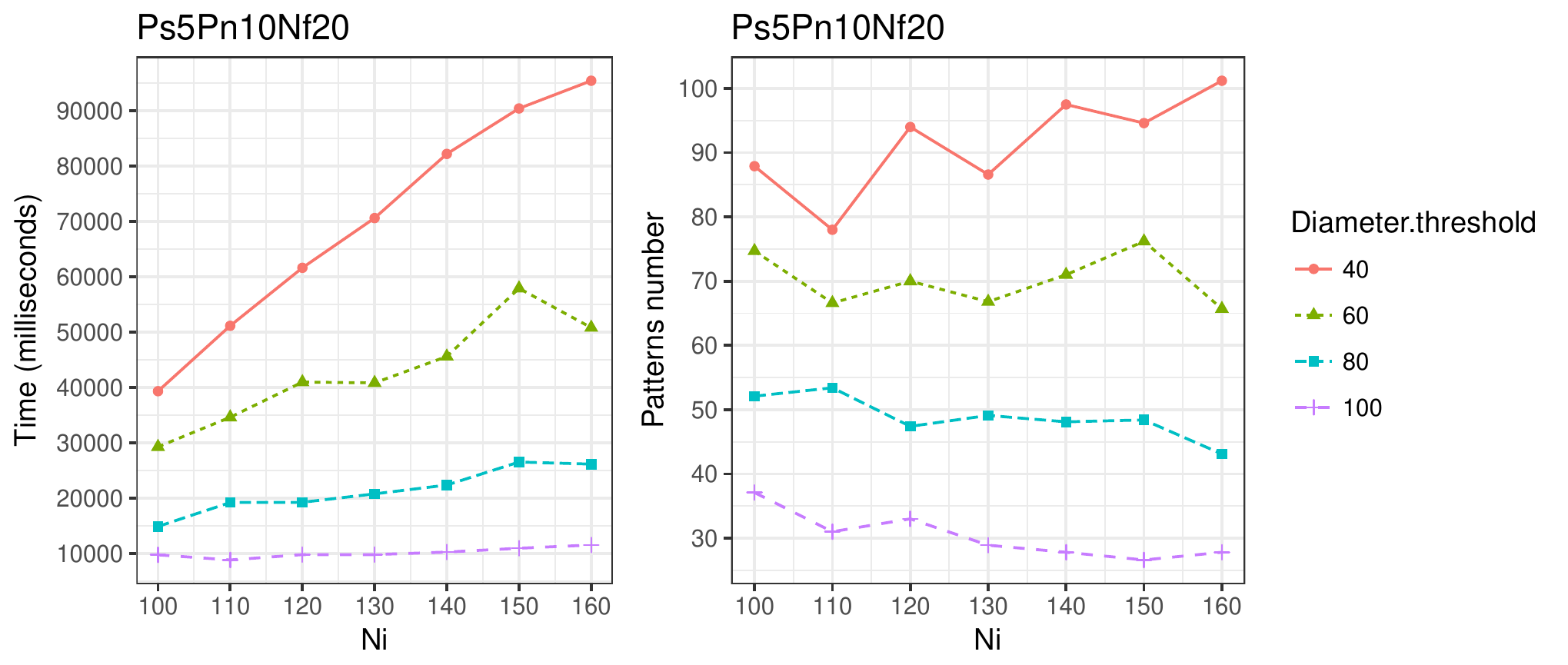}
	\caption[]{Performance and number of discovered patterns with respect to number of event sequences}
	\label{Fig:9}
\end{figure}

\begin{figure}[h!t]
	\centering
	\includegraphics[width=0.98\linewidth]{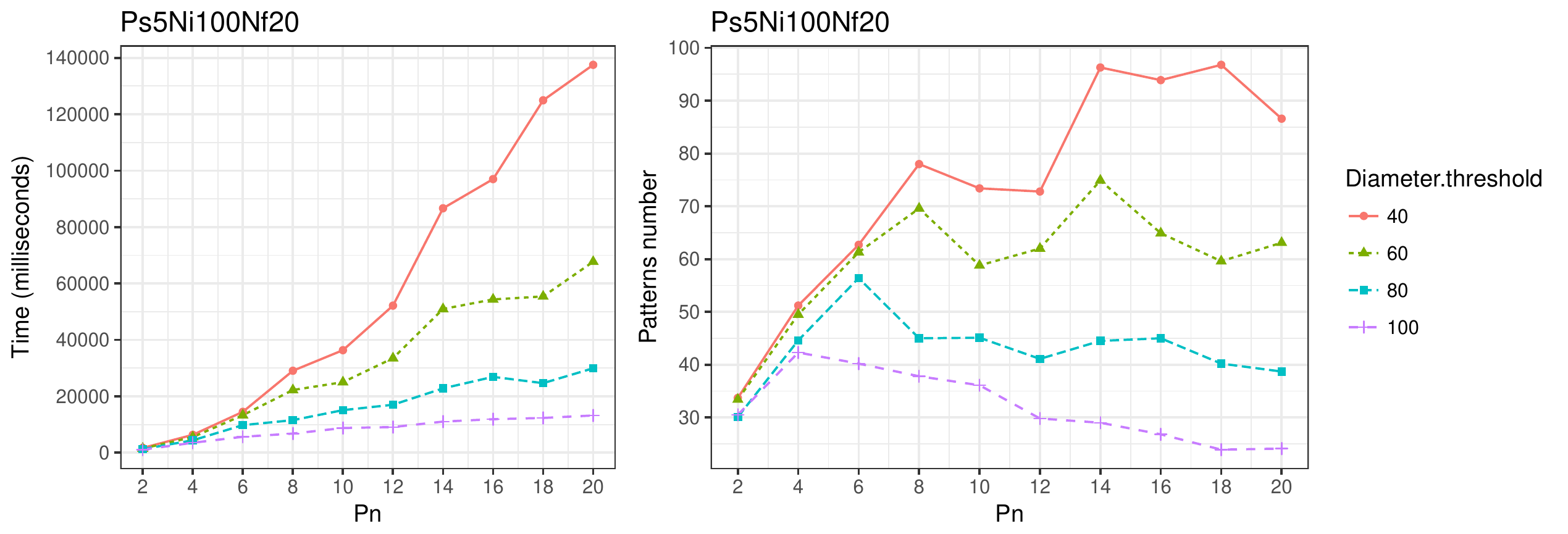}
	\caption[]{Performance and number of discovered patterns with respect to number of maximal patterns in dataset}
	\label{Fig:10}
\end{figure}

\begin{figure}[h!t]
	\centering
	\includegraphics[width=0.98\linewidth]{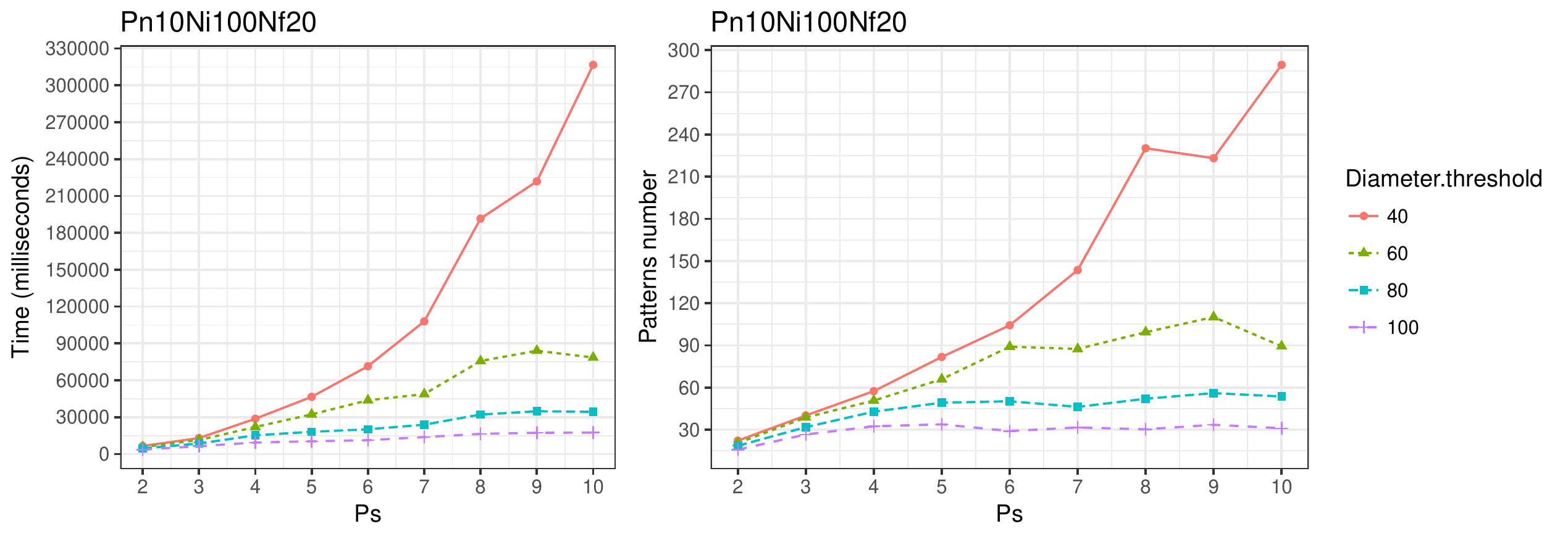}
	\caption[]{Performance and number of discovered patterns with respect to maximal patterns length}
	\label{Fig:11}
\end{figure}

We present here the results obtained for the Micro-ST-Miner algorithm and datasets described in Section~\ref{Sec:ExperimentalResults}. We set the sequence index threshold to constant $ 1 $. In Fig.~\ref{Fig:9},~\ref{Fig:10},~\ref{Fig:11} numbers $ 40, 60, 80, 100 $ are diameter thresholds for our Micro-ST-Miner algorithm. Additionally,  ST-Miner algorithm was not able to obtain any results for generated datasets and sequence index threshold $ 1 $ in reasonable time.

Results presented in Fig.~\ref{Fig:9},\ref{Fig:10},\ref{Fig:11} show improvements of our Micro-ST-Miner algorithm over ST-Miner algorithm presented in \cite{ref1284:Huang2008}. The execution times obtained by our approach are much better than obtained by ST-Miner for the same types of datasets.

\subsection{Results on Real Data}

In our experiments, we used weather data provided by ECAD project \cite{ref1284:Besselaar2011,ref1284:Haylock2008}. The dataset contains the following event types: Temperature, Pressure and Percipitation provided in the gridded form. The resolution of each grid is 0.5 deg. and contains daily observations for years 1950-2016. The example of visualization may be found on the web page \cite{ref1284:EOBSdata}. 

From the original dataset we extract abnormal events by taking for each event type observations from 0.01 quantile and with values greater than 0.99 quantile. This results in the following event types: \textit{High Temperature}, \textit{Low Temperature}, \textit{High Pressure}, \textit{Low Pressure}, \textit{High Percipitation} (in our experiments we omit \textit{Low Percipitation} event type). The resulting dataset contains 70922 instances. For the microclustering step we apply Algorithm~\ref{Alg:5} with \textit{K} values equal to: $ 20, 40, 60, 80, 100 $. Spatial and temporal diameters of microclusters are normalized and integrated into one diameter threshold as it was mentioned in Section~\ref{Sec:MircoClust}. We set spatial diameter threshold to 2.5 deg. and temporal diameter threshold 30 days.

\begin{table}[h!t]
	{\scriptsize 
		\caption{Calculation time (in milliseconds), microclustering index size and compression ratio after applying microclustering algorithm for real dataset with number of instances equal to 70922}
		\begin{tabularx}{\textwidth}{cXXXXX}
			\toprule
			& \multicolumn{5}{c}{K} \\
			\midrule \midrule 
			& 20 & 40 & 60 & 80 & 100  \\
			\cmidrule{2-6} 
			\addlinespace 
			Microclustering index size & 5236 & 2618 & 1759 & 1334 & 1081 \\
			\midrule \addlinespace 
			Compression ratio & 13.5451 & 27.0901 & 40.3195 & 53.1649 & 65.6078 \\
			\midrule \addlinespace 
			Microclustering time & 55657 & 42106  & 45245 & 40015 & 38898 \\			
			\bottomrule
		\end{tabularx}
	}
	\label{Table:7}	
\end{table} 

The set of possible patterns is as follows: \textit{High Percipitation} $ \rightarrow $ \textit{High Temperature}, \textit{Low Pressure} $ \rightarrow $ \textit{High Percipitation}, \textit{Low Temperature} $ \rightarrow $ \textit{Low Pressure} $ \rightarrow $ \textit{High Percipitation}. All patterns have been discovered using our Micro-ST-Miner algorithm with sequence index threshold set to 1.

\section{Conclusions}

In the paper, we proposed modification of recently proposed notions and algorithms for the problem of discovering sequential patterns from event-based spatiotemporal data. The proposed method starts with computing microclustering index for a given dataset and appropriately redefining notions of neighborhood, density and density ratio. We offered our algorithm for the microclustering step and algorithms for computing significant sequential patterns using the reduced dataset. In summary, we have the following most important conclusions:

\begin{enumerate}
	\item Results obtained for real dataset show the usefulness of proposed approach. However, future studies should focus on the more broad study of proposed approach using real datasets. 
	\item Presented experiments show the usefulness of the proposed algorithm for generated datasets. For generated datasets, ST-Miner was not able to obtain results in any reasonable time. Proposed Micro-ST-Miner algorithm allows to eliminate redundant and noise patterns from the dataset.   
\end{enumerate}

Further research problems in the described subject may focus on proposing methods for estimating or simulating appropriate value of the threshold of a pattern significance. Additionally, it may be shown that the above introduced definitions of density and space $V_{N(e)}$ are inappropriate for some datasets. In particular, for scenarios where occurrences of instances of one type attract with significant delay occurrences of other events, definition of density is ineffective. Other possible research may propose alternative methods for discovering sequential patterns in spatiotemporal data adopting algorithms developed for mining association rules or considering the problem of efficient mining the $K$ most important patterns. 

\textbf{Acknowledgements.} We acknowledge the E-OBS dataset from the EU-FP6 project ENSEMBLES (http://ensembles-eu.metoffice.com) and the data providers in the ECA\&D project (http://www.ecad.eu)"
"Haylock, M.R., N. Hofstra, A.M.G. Klein Tank, E.J. Klok, P.D. Jones, M. New. 2008: A European daily high-resolution gridded dataset of surface temperature and precipitation. J. Geophys. Res (Atmospheres), 113, D20119, doi:10.1029/2008JD10201 \cite{ref1284:Haylock2008}

We acknowledge the E-OBS dataset from the EU-FP6 project ENSEMBLES (http://ensembles-eu.metoffice.com) and the data providers in the ECA\&D project (http://www.ecad.eu)"
"van den Besselaar, E.J.M., M.R. Haylock, G. van der Schrier and A.M.G. Klein Tank. 2011: A European Daily High-resolution Observational Gridded Data set of Sea Level Pressure. J. Geophys. Res., 116, D11110, doi:10.1029/2010JD015468 \cite{ref1284:Besselaar2011}

We would like to kindly thank Professor Marzena Kryszkiewicz for her helpful comments and discussion during preparing the article.

\bibliographystyle{splncs03}
\bibliography{bib}  

\end{document}